# Development and validation of a conceptual multiple-choice survey instrument to assess student understanding of introductory thermodynamics

Mary Jane Brundage and Chandralekha Singh
Department of Physics and Astronomy, University of Pittsburgh, Pittsburgh, PA 15260

**Abstract.**
We discuss the development and validation of the long version of a conceptual multiple-choice survey instrument called the Survey of Thermodynamic Processes and First and Second Laws-Long (STPFaSL-Long) suitable for introductory physics courses. This version of the survey instrument is a longer version of the original shorter version developed and validated earlier. The 19 contexts including the exact wording of all of the problem situations posed in the two versions of the survey instrument are identical and the difference between the long and short versions of the instrument is only in the multiple-choice options. In particular, in the longer version of the survey instrument, there are no alternative conceptions explicitly embedded in the four multiple-choice options students choose from and the questions asked in a given context in one item of the shorter survey instrument were split into several items focusing, e.g., on different thermodynamic variables. After the development and validation of the longer version of the survey instrument, the final version was administered in 12 different in-person classes (four different institutions) in which students answered the questions in-class on paper scantron forms with the instructor as the proctor and 12 different in-person classes (five different institutions) in which students answered the questions online on Qualtrics within a two-hour period. This longer version of the survey instrument was administered to introductory physics students in various traditionally taught calculus-based and algebra-based classes before and after traditional lecture-based instruction in relevant concepts. It was also administered to upper-level undergraduates majoring in physics and Ph.D. students taught traditionally for bench marking purposes and for concurrent validity, which involved comparing advanced students' performance with those of introductory students for whom the survey is intended. Similar to the shorter version, we find that although the longer version of the survey instrument focuses on thermodynamics concepts covered in introductory courses, it is challenging even for advanced students. A comparison with the baseline data on the longer version of the validated survey instrument presented here can help instructors evaluate the effectiveness of innovative pedagogies designed to help students develop a solid grasp of these concepts.

## I. INTRODUCTION AND BACKGROUND

### A. Multiple Choice Surveys

Central goals of college introductory physics courses for life science, physical science and engineering majors include helping all students develop functional understanding of physics and learn effective problem solving and reasoning skills [1-10]. Validated conceptual multiple-choice physics survey instruments administered before and after instruction in relevant concepts can be useful tools to gauge the effectiveness of curricula and pedagogies in promoting robust conceptual understanding. When compared to free-response problems, multiple-choice problems can be graded efficiently and results are easier to analyze statistically for different instructional methods and/or student populations. However, multiple-choice problems also have some drawbacks. For example, students' selection of the correct answers does not necessarily reflect understanding of the underlying concepts. Also, students cannot be given partial credit for their responses.

Multiple-choice survey instruments have been used as one tool to evaluate whether research-based instructional strategies are successful in significantly improving students' conceptual understanding of these concepts [11-14]. Apart from the Force Concept Inventory, the most well-known survey, other conceptual survey instruments at the introductory physics level in mechanics and electricity and magnetism have also been developed, including survey instruments for kinematics represented graphically [15], energy and momentum [16], rotational and rolling motion [17, 18], electricity and magnetism [19-23], circuits [24] and Gauss's law [25, 26]. In thermodynamics, existing conceptual survey instruments include: 1) Heat and Temperature Conceptual Evaluation (HTCE) [27] that focuses on temperature, phase change, heat transfer, thermal properties of materials; 2) Thermal Concept Evaluation (TCE) [28, 29] that also focuses on similar concepts to HTCE; 3) Thermal Concept Survey (TCS) [30] that focuses on temperature, heat transfer, ideal gas law, first law of thermodynamics, phase change, and thermal properties of materials; 4) Thermodynamics Concept Inventory (TCI) [31] that focuses on concepts in engineering thermodynamics courses; and 5) Thermal and Transport Concept Inventory: Thermodynamics (TTCI:T) [32] that also focuses on concepts in engineering thermodynamics courses. In addition, recently a multiple response survey has been developed for upper-level thermal physics [33].

Despite the availability of the five conceptual survey instruments on introductory thermodynamics, there is a lack of research-validated survey instrument that focuses on the basic concepts related to thermodynamic processes and the first and second laws covered in introductory physics courses. Therefore, we earlier developed and validated [34, 35] a 33-item conceptual multiple-choice survey instrument on these concepts called the Survey of Thermodynamic Processes and First and Second Laws (STPFaSL-Short). Here we discuss the development and validation [36, 37] of a version of the survey instrument called STPFaSL-Long that is a longer version of the shorter survey developed and validated earlier. The survey instrument can be accessed from PhysPort and is included in Ref. [36] in the supplementary materials section. The 19 contexts—including the exact wording of all of the problem situations posed in the short and long versions of the STPFaSL survey instrument—are identical, and the difference between the long and short versions of the instrument is only in the multiple-choice options. In particular, in the longer version of the survey, there are no alternative conceptions explicitly embedded in the four multiple-choice options students choose from and the questions asked in a given context in one item of the shorter version of the survey were split into several items focusing, e.g., on different thermodynamic variables. We note that the overlap of the STPFaSL content with the HTCE and TCE is minimal. Moreover, although there is overlap between TCI, TTCI:T and STPFaSL concepts, contexts used in the TCI and TTCI:T are engineering focused and, therefore, these surveys are unlikely to be used by introductory physics instructors. Finally, the TCS is for introductory physics courses and covers some content common with STPFaSL but the TCS is a much broader survey and has a major emphasis on temperature, the ideal gas law, phase change and thermal properties of materials, content that are not explicitly the focus of the short and long versions of the STPFaSL instrument.

### B. Inspiration from prior investigations on student understanding of thermodynamics

Prior research has not only focused on the development and validation of multiple-choice surveys to investigate students' conceptual understanding of various thermodynamic concepts, but many investigations have focused on student understanding of thermodynamics without using multiple-choice surveys [38-66]. Some of these investigations use conceptual problems to probe student understanding and ask students to explain their reasoning. These investigations were invaluable in the development of both versions, STPFaSL-Short instrument [34, 35] and the STPFaSL-Long instrument discussed here. For brevity, below, we only give a few examples of studies that were used as a guide and from which open-ended questions were used in the earlier stages of the development of the multiple-choice questions for both versions of the instrument discussed here.

Loverude et al. [39] investigated student understanding of the first law of thermodynamics in the context of how students relate work to the adiabatic compression of an ideal gas. For example, in one problem used to investigate student understanding in their study, students were asked to consider a cylindrical pump (diagram was provided) containing one mole of an ideal gas. The piston fit tightly so that no gas could escape. Students were asked to consider friction as being negligible between piston and the cylinder. The piston was thermally isolated from the surroundings. In one version of the problem, students were asked what will happen to the temperature of the gas and why if the piston is quickly pressed inward. Another type of problem posed to students in the same research involved providing a cyclic process on a PV diagram in which parts of the cyclic process were isothermal, isobaric and isochoric. Students were asked whether the work done in the entire cycle was positive, negative or zero and to explain their reasoning.

Another investigation of students' reasoning about heat, work and the first law of thermodynamics in an introductory calculus-based physics course by Meltzer [40] asked several conceptual problems some of which involved PV diagrams. For example, one problem in this study involved two different processes represented on a PV diagram that started at the same point and ended at the same point. Students were asked to compare the work done by the gas and the heat absorbed by the gas in the two processes and explain their reasoning for their answers.

In another investigation focusing on student understanding of the ideal gas law using a macroscopic perspective, Kautz et al. [41] asked several conceptual problems. For example, in one problem in which a diagram was provided, three identical cylinders are filled with unknown quantities of ideal gases. The cylinders are closed with identical frictionless pistons. Cylinders A and B are in thermal equilibrium with the room at 200°C, and cylinder C is kept at a temperature of 800°C. The students were asked whether the pressure of the nitrogen gas in cylinder A is greater than, less than, or equal to the pressure of the hydrogen gas in cylinder B, and whether the pressure of the hydrogen gas in cylinder B is greater than, less than, or equal to the pressure of the hydrogen gas in cylinder C. Students were asked to explain their reasoning.

Another investigation by Cochran et al. [42] focused on student conceptual understanding of heat engines and the second law of thermodynamics. For example, in one question students were provided the diagram of a proposed heat engine (including the temperatures of the hot and cold reservoirs, the heat absorbed from the hot reservoir and the heat flow to the cold reservoir as well as the work done in one cycle) and asked if the device as shown could function and why. In another investigation, Bucy et al. [43] focused on student understanding of entropy in the context of comparison of ideal gas processes. For example, students were asked to compare the change in entropy of an ideal gas in an isothermal expansion with a free expansion in a vacuum, and also to explain whether the change in entropy of the gas in each case is positive, negative or zero in each case, and why. In another investigation by Christensen et al. [44], students' ideas regarding entropy and the second law of thermodynamics in an introductory physics course were studied. They found that students struggled in distinguishing between entropy of the system and the surroundings and had great difficulty with spontaneous processes. Another investigation by Smith et al. [45] focused on student difficulties with concepts related to entropy, heat engines and the Carnot Cycle and how student understanding can be improved.

### C. Rationale for developing STPFaSL-Long

Before we discuss the development and validation of the STPFaSL-Long instrument, we discuss the rationale for developing this long version when STPFaSL-Short version exists [34, 35]. In particular, all the scenarios for the 19 contexts and the wording of the questions before the options students select are identical in both of the versions of the survey instrument related to thermodynamic processes and the first and second laws covered in introductory physics courses. The difference between the long and short versions of the instrument is only in the multiple-choice options.

The rationale for developing the STPFaSL-Long version of the instrument came from discussions with some thermodynamics instructors after the STPFaSL-Short version was developed and validated. Some instructors noted that they prefer to give surveys in which no alternative conceptions of students are embedded in the questions since they did not want students to be misled to an incorrect answer by gravitating to different alternative conceptions. They felt that in the absence of the alternative conceptions, their students may perform significantly better because they would think about each option more carefully instead of being misled by alternative conceptions embedded in some of the questions in the short version of the instrument. Moreover, in the short version, students were often asked about more than one thermodynamic variable in a single question. Since there are only five choices students can select from for each question, they were only provided those choices in the multiple-choice questions that students had selected with highest frequency in the written open-ended administration of these questions and individual interviews during the development of the short version of the instrument. However, some instructors noted that they would prefer to use a version in which all possible choices for each thermodynamic variable were provided in each situation, e.g., the internal energy of a gas in a given situation increases, decreases, remains the same or there is not enough information.

Therefore, based upon the feedback from these instructors, we developed and validated the longer version of the instrument, in which there are no alternative conceptions explicitly embedded in the four multiple-choice options students are asked to choose from. Moreover, the questions asked in a given context in one item of the shorter version of the instrument were split into several items focusing, e.g., on different thermodynamic variables.

To give a concrete example, item 1 in the short version of the instrument asks students to choose the correct statement about the change in entropy of the gas that undergoes a reversible adiabatic expansion with incorrect options involving alternative conceptions such as "Entropy of the gas increases because the gas expands" or "Entropy of the gas remains constant because the entropy of the gas does not change for a reversible process". On the other hand, the corresponding item in long version of the instrument does not have alternative conceptions embedded in the options. It asks students to choose the correct statement about the change in entropy of the gas that undergoes a reversible adiabatic expansion with options such as entropy of the gas increases, decreases, remains constant and there is not enough information.

To give another concrete example, item 2 in the short version asks which one of the following statements must be true for the gas that undergoes a reversible adiabatic expansion process:

a. The internal energy must decrease, and the work done by the gas must be positive.
b. The internal energy must decrease, and the work done by the gas must be negative.
c. The internal energy must increase, and the work done by the gas must be positive.
d. The internal energy must increase, and the work done by the gas must be negative.
e. None of the above.

Since not all possible choices can be provided for internal energy and work done by the gas in the short version (e.g., there is no option with one or both of them remaining constant), the long version of the instrument splits this item into two separate items, one of which asks about internal energy and the other about the work done by the gas with each question providing all possible options (e.g., the internal energy must decrease, increase, remain constant and there is not enough information and the work done by the gas must be positive, negative, zero or there is not enough information).

Based upon discussions with instructors, we believe that different instructors may prefer one version or another based upon the features of the two versions they value. For example, some instructors may prefer the short version which has alternative conceptions embedded in some of the items and in which an item can ask about two thermodynamic variables with restricted options (as in the example with internal energy and work discussed above) since those restricted options correspond to the most common choices students selected in the written open-ended questions and individual interviews. However, as noted, other instructors may prefer the long version with no alternative conceptions and an item being restricted to only one thermodynamic variable with all possible options provided in the choices for that variable. We note that when we administered both the long and short versions of the instrument to the first-year first semester Ph.D. students in physics at an interval of approximately two weeks, the Pearson correlation coefficient between the two versions was 0.88, which is relatively high suggesting that both versions are measuring very similar things. Thus, instructors can use the version they prefer. Moreover, both versions take approximately the same amount of class time and can be administered in a 50-minute class period.

## II. STPFaSL INSTRUMENT DEVELOPMENT AND VALIDATION

The development and validation process of the two versions of the survey instrument, STPFaSL-Long and STPFaSL-Short [34, 35], were analogous to those for the earlier conceptual survey instruments developed by our group [16, 18, 21, 22, 25]. Our process is consistent with the established guidelines for test development and validation using the Classical Test Theory (CTT) [37]. According to the standards for multiple-choice survey instrument design, a high-quality test instrument should have five characteristics: reliability, validity, discrimination, good comparative data and suitability for the population [34, 35]. Moreover, the development and validation of a well-designed survey instrument is an iterative process that should involve recognizing the need for the survey instrument, formulating the test objectives and scope for measurement, constructing the test items, performing content validity and reliability checks, and distribution [34, 35].

Below, we describe the development, validation and administration of the STPFaSL-Long version of the instrument by first summarizing how the identical contexts for both the short and long versions were developed and validated, how the items involving various contexts in the short version were split into several items without changing the wording of the scenarios posed, and then presenting data from the administration of the final long version of the survey instrument. As noted earlier, the contexts are identical for the two versions of STPFaSL instrument and the difference between them is in the answer options students must choose from.

### A. Development of test blueprint, formulation of test objectives and scope

More details about the STPFaSL-Short version of the survey instrument development can be found in Ref. [34, 35] which is relevant for the development and validation of the long version discussed here. Summarizing the development common for both versions, we note that before developing the instrument, we first developed a test blueprint to provide a framework for deciding the desired test attributes [34, 35]. The test blueprint provided an outline and guided the development of the test items. The development of the test blueprint entailed formulating the need for the survey instrument, determining its scope, format and testing time of the test as well as determining the weights of different sub-topics consistent with the scope and objective of the test. The specificity of the test plan helped to determine the extent of content covered and the complexity of the questions. As noted in the introduction, despite the existence of several thermodynamics survey instruments at the introductory level [27-32], there is no research-validated survey instrument that focuses on the basics of thermodynamic processes and the first and second laws of thermodynamics covered in the introductory physics courses. Therefore, we developed and validated an instrument (with two versions: STPFaSL-Short and STPFaSL-Long) focusing on these content areas covered in introductory physics courses. With regard to the weights of different sub-topics consistent with the scope and objective of the test, we browsed over introductory physics textbooks, consulted with seven faculty members and looked at the kinds of questions they asked their students in homework, quizzes and exams before determining the 19 contexts.

    Both versions of the STPFaSL instrument have multiple-choice conceptual items on thermodynamic processes and the first and second laws covered in both calculus-based and algebra-based introductory physics courses, however, no calculus is required for students to complete the surveys. Both versions can be used to measure the effectiveness of traditional and/or research-based approaches for helping introductory students learn thermodynamics concepts covered in the survey for a group of students. Specifically, the survey instrument is designed to be a low stakes test to measure the effectiveness of instruction in helping students in a particular course develop a good grasp of the concepts covered and is not appropriate for high stakes testing. The survey instrument can be administered before and after instruction in relevant concepts to evaluate introductory physics students' understanding of these concepts and to evaluate whether innovative curricula and pedagogies are effective in reducing the difficulties. With regard to the testing time, the survey instrument (both short and long versions) is designed to be administered in one 50-minute class period although instructors should feel free to give extra time to their students as they deem appropriate. As noted, the reason the longer version of the STPFaSL does not take significantly more time is that all contexts and their wording are identical to the shorter version and each item in the longer version is about one thermodynamic variable (e.g., entropy, internal energy, heat transfer, work done by the system). Additionally, as noted, the answer choices in the long version do not explicitly include student alternative conceptions, but rather focus on how each of these individual variables change; answer choices are "increases," "decreases," "remains the same," "not enough information" or the questions in the longer version are framed as true/false questions. One exception is that in addition to the stem of the items, the four answer choices in item 36 in the longer version and item 19 in the shorter version are identical.

    We focused the survey content on thermodynamic processes and first and second laws that is basic enough that the survey instrument is appropriate for both algebra-based and calculus-based introductory physics courses in which these thermodynamics topics are covered. We also made sure that the survey instrument has questions at different levels of cognitive achievement [34, 35]. To formulate test objectives and scope pertaining to thermodynamic processes and first and second laws, the survey instrument development started by consulting with seven instructors who regularly teach calculus-based and algebra-based introductory physics courses in which these topics in thermodynamics are covered. We asked them about the goals and objectives they have when teaching these topics and what they want their students to be able to do after instruction in relevant concepts. In addition to perusing the coverage of these topics in several algebra-based and calculus-based introductory physics textbooks, we browsed over homework, quiz, and exam problems that these instructors in introductory algebra-based and calculus-based courses at a large research university had typically given to their students in the past before determining the test objective and scope of the test in terms of the actual content, and starting the design of the questions for the instrument. The preliminary distribution of questions from various topics was discussed and iterated several times, and finally agreed upon with seven introductory physics course instructors at a large research university.

    The selection of topics for the questions included consultation with 7 instructors who teach introductory thermodynamics (some of whom had also taught upper-level thermodynamics) about their goals and objectives and the types of conceptual and quantitative problems they expected their students in introductory physics courses to be able to solve after instruction. The wording of the questions took advantage of the existing literature regarding student difficulties in thermodynamics, input from students' written responses and interviews and input from physics instructors who teach these topics. In addition to leveraging the findings of prior research on students' understanding of these concepts [38-66], the process of administering some written open-ended questions and individual interviews with students at a large research university was helpful to develop the survey items [34, 35]. Moreover, as part of the development and validation of the survey, the concepts involved in the STPFaSL instrument, and the wording of the questions have been independently evaluated by four physics faculty members who regularly teach thermodynamics at the large research university (in addition to the feedback from members of the Physics Education Research or PER group at the large research university) and iterated many times until agreed upon. Moreover, two faculty members from

other universities who are experts in thermodynamics PER provided invaluable feedback several times to improve the quality of the survey questions.

As noted, the scope of both the short and long versions of the survey instrument is the same (identical 19 contexts and wording of the stem of the questions) except that the longer version splits each context in the shorter version into several items and each asks about only one thermodynamic variable with choices such as "increases," "decreases," "remains the same," or "not enough information," while some of the questions are framed as true/false questions. Thus, the longer version separates questions about different thermodynamic variables for the same context and scenario in order to make it easier for instructors to disentangle difficulties with different thermodynamic variables in a given scenario. In addition to separating questions about different thermodynamic variables for the same context and scenario, the four choices provided to students for each question in the longer version do not have any student alternative conceptions unlike the shorter version, which would appeal to instructors who prefer not to have items with alternative choices embedded in the choices provided.

As described in detail in Ref. [34, 35], we interviewed individual students using a think-aloud protocol at various stages of the survey instrument development to obtain a better understanding of students' reasoning processes when they were answering the free-response and multiple-choice questions. Fine-tuning of the instrument based upon statistical analysis using classical test theory (to be discussed in the next section for the longer version) was conducted on different iterations of both the short and long versions of the survey instrument as the items were being refined. Within this interview protocol, students were asked to talk aloud while they answered the questions so that the interviewer could understand their thought processes. Individual interviews with students during development of the survey instrument were useful for an in-depth understanding of the mechanisms underlying common student difficulties and to ensure that students interpreted the questions appropriately. Based upon the student feedback, the questions were refined and tweaked. During various stages of the development and validation process, 24 students in various algebra-based and calculus-based physics courses participated in the think-aloud interviews. Ten graduate students and undergraduates who had learned these concepts in an upper-level thermodynamics and statistical mechanics course were also interviewed while taking the shorter version (whose stems are identical to the longer version). In addition, 11 introductory physics students and 6 physics graduate students were interviewed while answering the STPFaSL-Long version of the instrument using think-aloud protocol to understand how they reasoned about different choices on each question on the longer version. The purpose of involving some advanced students in these interviews was to compare the thought processes and difficulties of the advanced students in these courses with introductory students for bench marking purposes. This type of bench marking has been valuable to illustrate growth of student understanding in prior research [64]. We found that students' reasoning difficulties across different levels are remarkably similar except in a few instances, e.g., advanced students were more facile at reasoning with PV diagrams than introductory students.

The final version of the STPFaSL-Long survey instrument has 78 multiple-choice items including 22 true/false questions and each question has one correct choice and three incorrect choices. To aid instructors who administer STPFaSL-Long in their classes, Figs. 1a-1c in Appendix A classify the broad categories of topics covered in each of the 78 items of the STPFaSL-Long survey instrument as Processes, Systems, Quantities & Relations, Representation, the First Law of Thermodynamics, and the Second Law of Thermodynamics. The Processes category includes items which require understanding of thermodynamic constraints such as whether a process is reversible, isothermal, isobaric or adiabatic. Also included are problems involving irreversible and cyclic processes. We note that these different processes are not necessarily exclusive; e.g., one can consider an isothermal reversible process. The Systems category includes items involving knowledge of the distinction between a system and the universe, items involving subsystems or an isolated system. The Systems category also includes items in which a student could make progress by making use of the fact that the system is an ideal gas (e.g., for an ideal gas, the internal energy and temperature have a simple relationship which can be used to solve a problem). Quantities and Relations includes survey items specific to a quantity such as internal energy, work, heat, entropy, and their quantitative relationships. For example, the relationship between work and the area under the curve on a PV diagram is tested in several problems. The Representation category includes items in which a process is represented on a PV diagram. Finally, the last two categories include items requiring the First Law and Second Law of Thermodynamics. We classified questions about heat engines into the Second Law of Thermodynamics category (although heat engines involve both the first and second laws) due to the particular focus of these.

### B. Validation of the survey instrument

While developing and validating the STPFaSL survey instrument (both short and long versions), we paid particular attention to the issues of reliability and validity [37, 67-70]. Test reliability refers to the internal consistency of the test or relative degree of consistency between the scores if an individual immediately repeats the test, and validity refers to the appropriateness of interpreting the test scores [37, 67-70]. We note that the STPFaSL instrument is appropriate for making interpretations about the effectiveness of instruction in relevant concepts in a particular course and it is not supposed to be used for high stakes testing of individual students.

Face validity of the test refers to whether it is a good measure of students' understanding of the concepts measured by the test and content validity of a test refers to whether the experts in the discipline who review the test agree that the items in the test adequately cover all aspects of the construct being measured [37]. In the earlier subsections of section II, we already discussed face validity and content validity including how expert and student involvement (via open-ended questions and individual interviews) was used for this purpose. Also, although the survey instrument focuses on concepts that are typically covered in introductory thermodynamics and is appropriate for introductory physics courses, it was also administered to undergraduates in upper-level thermodynamics and statistical mechanics courses in which these concepts are generally covered and to first-year first-semester physics Ph.D. students to obtain base line data and to check for one form of criterion validity or concurrent validity [71, 72]. In particular, we discuss the concurrent validity of the STPFaSL-Long survey instrument using comparison between introductory and advanced students' performance (e.g., whether introductory students outperformed advanced students on the survey). Since the quantitative measures validating the STPFaSL-Short instrument were described earlier [34, 35], below we focus only on the STPFaSL-Long instrument in terms of the quantitative measures used in the classical test theory for a reliable survey instrument including item analysis (using item difficulty and point biserial coefficient) and KR-20 [37, 73, 74].

### C. Proctored and unproctored administration of STPFaSL-Long

After the development and validation of STPFaSL-Long, data were collected using the final version administered in various classes. This includes 12 different in-person classes (four different institutions) in which students answered the questions in-class on paper scantron forms (proctored administration) and 12 other different in-person classes (five different institutions) in which students answered the questions online on Qualtrics within a two-hour period from the time they started the survey (unproctored administration). Even though all classes were taught in-person, the reason for administering STPFaSL-Long unproctored via Qualtrics in half of the classes is that many course instructors were reluctant to spend an entire class period to administer it. Considering physics instructors often feel that they must cover a lot of content and cannot spare a class period to administer a survey, this issue of online administration even in in-person classes is commonly encountered in physics classes [75]. Since the data from different institutions for the same type of course (e.g., calculus-based introductory physics course) are similar, average combined data from different institutions for the same course type are presented here. All data presented here are from primarily traditional lecture-based physics courses (in-class administration data are in Appendix B, Qualtrics data for students who took the survey online are in Appendix C and the STPFaSL-Long survey instrument is in the supplementary materials) so that instructors in courses covering the same concepts but using innovative pedagogies can compare their students' performance with those provided here to gauge the relative effectiveness of their instructional design and pedagogies. We combined the upper-level undergraduates and first-year first-semester physics Ph.D. students since the first-year graduate students only had thermodynamics instruction in their upper-level undergraduate courses.

In the in-person administration of the full 78-item survey instrument, students were provided the entire class period (typically 50 minutes) but in the online administration of the survey instrument using Qualtrics, students were provided two-hours from the time they started the survey in one sitting in a given week after instruction in relevant concepts. In consultation with several instructors, the reason students were provided two-hours for the online unproctored administration via Qualtrics is that since these students are not taking the survey in class, they may get interrupted by various things including visitors and phones but we did not want to give students unlimited time to limit the opportunity to look up answers or consult with others (note that the unproctored Qualtrics version explicitly noted that students were not supposed to consult any resources including books, notes, internet, friends). Moreover, STPFaSL-Long was administered before instruction (as a pretest) in relevant concepts to investigate student understanding before instruction in some of the introductory physics courses at one university only because other instructors did not think it was necessary and were reluctant to administer the survey both before and after instruction in their classes. For all pre-test data and some of the in-person administered introductory posttest data, students were given only the first 48 items or the last 52 items of the 78-item survey. In all undergraduate courses, students were given extra credit incentives to take the survey consistent with suggestions of Effective Practices for Physics Programs Guide section on How to Select and Use Various Assessment Methods in your Program [76]. In particular, if students are not given any grade incentives, some students do not take the survey seriously, but the survey should be a low stakes assessment consistent with our goals.

## D. Overall performance, reliability, item difficulty and point biserial coefficient

After the development and validation of STPFaSL-Long instrument, as noted, the final version was administered as proctored assessment in 12 different classes (four different institutions) in which students answered the questions in-class on paper scantron forms and as unproctored assessment in 12 different classes (five different institutions) in which students answered the questions online on Qualtrics within a two-hour time-window. All 24 classes were taught in-person. The survey instrument was administered to introductory physics students in various traditionally taught calculus-based and algebra-based classes after traditional lecture-based instruction in relevant concepts. It was also administered to upper-level undergraduates majoring in physics, taught traditionally, and Ph.D. students for benchmarking purposes and for one type of criterion validity (concurrent validity), which involved comparing their performance with those of introductory students for whom the survey is intended. We find that although the survey instrument focuses on thermodynamics concepts covered in introductory courses, it is challenging even for advanced students. Moreover, in some introductory courses at one university, as noted, the pretest was administered before students learned about thermodynamics in that course and the posttest was administered after instruction in relevant concepts. We note that some introductory physics instructors teach thermodynamics at the end of the first semester introductory physics course while others teach it at the beginning or in the middle of another second semester course. We did not find any trends based upon these, so all data are included for a particular group of students.

While students who took the survey online via Qualtrics were told that it was a survey in which they were not supposed to consult any resources, we did not have any method for verifying that. On the other hand, the in-class administration was proctored by the instructor of the course. Due to proctored and unproctored nature of the administration, we feel that these data are best kept separate so that instructors using similar mode of administration for a particular group of students can compare their course's performance with the averages provided here.

Table I shows in-class administration data pertaining to average student performance on the STPFaSL-Long instrument from introductory calculus-based and algebra-based courses for matched and unmatched students on pretest (before instruction) and posttest (after instruction) from a university where the pretest was also administered. For calculus-based courses, some of the instructors only administered the pretest or posttest, but not both. The matched and unmatched data in Table 1 are only for those instructors who administered both pretest and posttest. This table shows that the matched and unmatched data are very similar. Therefore, in the rest of the paper, we include all data available for a particular group. Moreover, we conducted a *t*-test and found that the scores of the calculus-based and algebra-based classes (both for the pretest and posttest) are not statistically significantly different. However, we have kept these scores separate since many instructors/researchers are often only interested in data from calculus-based physics classes or algebra-based physics classes.

TABLE I: Data showing average student performance on the STPFaSL-Long instrument from introductory calculus-based and algebra-based courses for matched and unmatched students on pretest (before instruction) and posttest (after instruction) from a university where the pretest was also administered.

|  | Pretest | | Post-Test | |
| --- | --- | --- | --- | --- |
|  | Matched | Unmatched | Matched | Unmatched |
| Intro calculus | 52.1% <br> N = 214 | 51.6% <br> N = 266 | 56.6% <br> N = 214 | 56.7% <br> N = 241 |
| Intro algebra | 51.0% <br> N = 315 | 50.9% <br> N = 371 | 57.1% <br> N = 315 | 57.0% <br> N = 388 |

One way to measure reliability of the test instrument is to prepare an ensemble of identical students, administer the test instrument to them, and analyze the resulting distribution on each item and overall scores. Since this is generally impractical, instead, a method is devised to use subsets of the test itself and consider the correlation between different subsets. The Kuder-Richardson reliability index or KR-20 reliability index [37, 73, 74], which is a measure of the self-consistency of the entire test instrument, can take a value between 0 and 1 (it divides the full instrument into subsets and the consistency between the scores on different subsets is estimated). If guessing is high, KR-20 will be low. Table II shows the number of students in each group averaged across similar classes to whom the final version of the survey was administered, as well as the average performance of different groups on the entire survey instrument for in-class and online administrations using Qualtrics; KR-20 for the post-tests for each group that took the entire survey are also shown. The KR-20 values for all groups in Table II are reasonable for predictions about a group of students [38, 74, 75]. We note that the written student data are from 12 different in-person courses from four different large public institutions; students completed the survey in class on Scantrons during a 50-minute class period if they took the entire survey since some students were administered only the first 48 or last 52 questions. In particular, not all introductory students answered all survey questions, and some introductory students were given only the first 48 or last 52 questions during recitation sessions to ensure split-test reliability. On the post-test (pre-test), out of the 492 (753) Int-calc students, 168 (505) were given the first 48 questions, 73 (248) were given the last 52 questions, and 251 (0) were given the full survey. On the post-test (pre-test), out of the 550 (371) Int-alg students, 170 (173) were given the first 48 questions, 218 (198) were given the last 52 questions, and 162 (0) were given the full survey.

The item difficulty of each multiple-choice question on the STPFaSL-Long instrument is simply the percentage of students who correctly answered the question, i.e., it is the average score on a particular item. Results in Table III in Appendix B show not only the item difficulty of each item on the instrument but also the prevalence of different incorrect choices for each item for each group for in-class administration. The corresponding data for online administration using Qualtrics are shown in Table IV in Appendix C.

TABLE II. The average performance and standard deviation (SD) of different groups on the STPFaSL-Long, the number (N) of students who participated in the survey in each group as well as KR-20 for posttests when students took the entire survey. "Upper Post" consists of advanced undergraduate students who had learned the relevant concepts in an upper-level thermodynamics and statistical mechanics course and first-year physics Ph.D. students in their first semester of the Ph.D. program who had not taken the graduate level thermodynamics and statistical mechanics course at their institution. Pretest (Pre) was administered before students learned relevant concepts in the course and posttest (Post) was administered after relevant instruction in the calculus-based and algebra-based introductory physics courses. For in-person administration of the survey for introductory students only, some students were given the entire 78-item survey and this is shown by (78). Some students were given only the first 48 items or the last 52 items of the 78-item survey, and these groups are represented with (48) and (52), respectively.

| Level | In-person Pre-test (48) | In-person Pre-test (52) | In-person Posttest (78) | In-person Posttest (48) | In-person Posttest (52) | Online Posttest (78) |
|---|---|---|---|---|---|---|
| Upper | … | … | 76% | … | … | 68% |
| | … | … | N = 89 | … | … | N = 179 |
| | … | … | SD = 14% | … | … | SD = 17% |
| | | | KR-20 = 0.90 | | | KR-20 = 0.92 |
| Int-calc | 52% | 52% | 58% | 58% | 55% | 70% |
| | N = 505 | N = 248 | N = 251 | N = 168 | N = 73 | N = 376 |
| | SD = 9% | SD = 10% | SD = 11% | SD = 11% | SD = 12% | SD = 19% |
| | | | KR-20 = 0.80 | | | KR-20 = 0.94 |
| Int-alg | 51% | 51% | 52% | 55% | 58% | 60% |
| | N = 173 | N = 198 | N = 162 | N = 170 | N = 218 | N = 450 |
| | SD = 9% | SD = 10% | SD = 11% | SD = 11% | SD = 13% | SD = 20% |
| | | | KR-20 = 0.77 | | | KR-20 = 0.94 |
| Average KR-20 for All Posttests (78) | | | 0.87 | | | 0.94 |

The Point Biserial Coefficient or PBC is designed to measure how well a given item predicts the overall score on a test. It is defined as the correlation coefficient between the score for a given item and the overall score. The PBC can take on values between -1 and 1; a negative value indicates that otherwise high-performing students score poorly on this item, and otherwise low-performing students do well on the item. The point biserial coefficients are shown in Figure 2 for in-person administration and Figure 3 for online administration in Appendix D. A commonly used criterion [37, 73, 74] states that it is desirable for this measure to be greater than or equal to 0.2, which is exceeded for 71 of the 78 items on the STPFaSL for the in-person implementation (and is lower than 0.2 only for one question for the online implementation at 0.16). Five of these questions with a PBC less than 0.2 focus on entropy in which the advanced students did not necessarily do better than the introductory group.

### E. Note on items with point biserial coefficient below the threshold

Five of the seven questions below the threshold for the in-class administration (items 5, 8, 24, 48 and 64) are related to entropy on which advanced students' average performance is comparable to introductory physics students' average performance. Our analysis shows that many advanced students performed poorly on these questions because they incorrectly thought that the entropy of a system always increases (e.g., in one cycle of a cyclic process). These findings are consistent with Meltzer's findings [52]. The other two items with PBC below the threshold are item 36 (basics of Carnot engine) and item 70 (magnitude of work done in an adiabatic process is the same as magnitude of heat transfer in a constant volume process if the change in internal energy is same in both processes). On item 36, the average scores of all groups are at or below 40%, and for item 70, it is at or below 30% which implies that these items are extremely difficult for students at all levels. The poor performance of all groups reduced the PBC. We believe that the issues covered in these two items are important issues that instructors should focus on to help students at all levels learn them better so we keep these items.

### F. Reliability via splitting the survey and administering each half to different students in introductory classes

We also tested reliability by splitting the survey instrument in two parts [37] without splitting related questions involving a given scenario, with 48 questions in one part and 52 questions in another part; 22 items were common between the two parts (these 22 items common items were the last 22 questions in the part with 48 questions and the first 22 questions in the part with 52 questions). At one university, we randomly administered to some students in introductory physics classes one part (48 questions) or the other part (52 questions) in-class. The performances of calculus-based introductory physics students who were administered the two version on the common 22 questions were 58.5% vs. 58.8%, which is similar, providing further evidence of reliability of the survey particularly because the order in which students were administered these questions was different, i.e., for some students these were the last 22 questions while for the others they were the first 22 questions. Also, the performances of these students on all 48 or 52 questions were 57.6% and 55.0%, respectively, showing that the first and second halves of the survey are relatively comparable in difficulty.

### G. Concurrent validity via administration to student groups at different levels

Earlier we discussed other forms of validity, e.g., face validity, content validity etc. Here we discuss one type of criterion validity or concurrent validity from administering the survey instrument to upper-level students and Ph.D. students (advanced students) [71, 72]. The criterion validity shows how well a test correlates with an established standard of comparison called a criterion [37]. One measure of this type of validity or concurrent validity can come from the expectation that introductory students will be outperformed by advanced students. As noted, a large number of students from introductory courses in which these concepts were covered were administered the final version of the survey instrument and advanced students in thermodynamic and statistical mechanics courses were administered the survey for the purposes of bench marking and concurrent validity of the instrument (see Table II).

The average data for calculus-based and algebra-based introductory students as well as advanced students tabulated in Table II show the expected trends for proctored in-class administration of the survey. In particular, advanced students with 76% average outperformed introductory physics students on the posttest (calculus-based courses averaged 57% and algebra-based courses averaged 55%). The average pretest performances of introductory physics groups were 52% and 51% (lower than posttest). These expected trends serve as a measure of concurrent validity although the STPFaSL survey is so difficult that none of the groups have stellar performance.

The data for the unproctored online administration (see Table II) show that advanced students obtained an average of 68% vs. 70% and 60%, respectively, for the calculus-based and algebra-based introductory physics courses after instruction in relevant concepts. Thus, introductory physics groups performed better in the online administration but the calculus-based introductory students performed exceptionally well comparable to the advanced students. The difference between the average scores for the proctored in-person and un-proctored Qualtrics administration for advanced students was -8%, for the algebra-based introductory physics students was 5% and for the calculus-based introductory physics students was 13%. In particular, in the online administration via Qualtrics, even though advanced students performed better than algebra-based physics post, calculus-based post averages are comparable to advanced student averages. We note that in the unproctored administration, the concurrent validity holds only for comparison between algebra-based introductory physics students and advanced students. It is not possible to pin-point the reasons for why calculus-based introductory students performed so well (comparable to advanced students), in the unproctored online administration of the survey. Introductory students may be more relaxed when taking the unproctored survey out of class which may be one reason that these groups on average performed better. However, we hypothesize that the better online performance may at least partly be due to a higher percentage of calculus-based introductory students consulting resources (they were asked not to consult with) in the un-proctored Qualtrics survey compared to the advanced students. In particular, while all students were told that the online survey was a closed book and closed notes test and students were not supposed to consult other resources, if some students in a particular group consulted external resources when answering the survey questions, that can affect the overall performance of the group [77, 78]. These data are still useful for what we can expect from an online administration of the survey to each group of students in which student response counted for extra credit [76]. In particular, it is still very useful to have the unproctored Qualtrics data since many instructors are only willing to administer these surveys online and they should be aware of these constraints when analyzing their own class data.

### H. Validity via measuring correlation between long and short versions

Some first-year first semester graduate students were administered both versions of the STPFaSL survey. The Pearson correlation between the short and long versions of the STPFaSL survey instrument for these students who were administered both versions of the survey at an interval of approximately two weeks is 0.88, which is a relatively high correlation showing that both the short and long versions measure very similar things and instructors can use whichever version they prefer. This high correlation with the already validated shorter version of the survey further demonstrates the validity of the longer version compared to the already validated shorter version [34, 35].

### I. A glance at student difficulties on the validated survey

Details about student difficulties found using the STPFaSL-Long instrument and comparison with STPFaSL-Short or prior studies are beyond the scope of this paper and will be presented elsewhere. However, we note that since the STPFaSL-Long instrument has been administered to a large number of students at different institutions for in-person and online formats, quantitative conclusions can be drawn about the prevalence of the many conceptual difficulties students have with these fundamental concepts in thermodynamics (see Table III for average performance of each group on each item for in-class administration in Appendix B and Table IV in Appendix C for online administration). Some of the conceptual difficulties displayed on the survey instrument include difficulty reasoning with multiple quantities simultaneously, difficulty in systematically applying various constraints (for an isothermal, adiabatic, isochoric, reversible, or irreversible process, isolated system, etc.), difficulty due to oversimplification of the first law and overgeneralization of the second law. As noted, many of these difficulties were inspired and incorporated in the survey instrument based upon those that have been documented (e.g., see Ref. [38-66]). Moreover, our findings with this validated survey instrument demonstrate the robustness of the previous findings, e.g., in Ref. [38-66], about student difficulties with these concepts in old contexts that have previously been studied and new contexts.

## III. SUMMARY

We developed, validated and administered the longer version of the STPFaSL instrument, which is a conceptual multiple-choice test focusing on thermodynamic processes and the first and second laws at the level covered in introductory physics courses. This survey instrument is a longer version of the shorter version developed and validated earlier in 2015. The 19 contexts including the exact wording of all of the problem situations posed in the two survey instruments are identical. The difference between the long and short versions of the instrument is only in the multiple-choice options. In particular, in the STPFaSL-Long survey, there are no alternative conceptions explicitly embedded in the four multiple-choice options students choose from and the questions asked in a given context in one item of the shorter survey were split into several items focusing, e.g., on different thermodynamic variables. The concepts related to thermodynamic processes and the first and second laws focusing on topics covered in an introductory physics course were challenging even for advanced students who were administered the survey instrument for obtaining baseline data and for evaluating content validity. The STPFaSL instrument is designed to measure the effectiveness of traditional and/or research-based approaches for helping introductory students learn thermodynamics concepts. The average individual scores on the survey instrument from traditionally taught classes at various institutions included in this study are low. We note that the average scores for other conceptual survey instruments for traditionally taught introductory classes are also low, e.g., for the Brief Electricity and Magnetism Assessment [19], the posttest scores for introductory students range from 23% to 45%, and for the Conceptual Survey of Electricity and Magnetism [20], the scores range from 25% to 47%. The low scores even after instruction indicate that the traditional instructional approach using lectures alone is ineffective in helping students learn these concepts. The instructors can choose the longer or shorter version of the STPFaSL survey instrument depending upon their preference to measure the effectiveness of instruction in these topics using a research-based pedagogy.

## ACKNOWLEDGMENTS
We thank David Meltzer and Mike Loverude for providing extensive feedback on several versions of the survey instrument. We also thank Jessica Clark for feedback on one version of the survey. We also thank David Meltzer and

Robert P. Devaty for helpful feedback on the manuscript. We thank all faculty members and students from all institutions who helped during the development and validation of the survey instrument and/or administered the final version to their classes.

## APPENDIX A: Relevant Topics for Each Question

Appendix A gives an example of relevant topics for each question.

| | Item # | 1 | 2 | 3 | 4 | 5 | 6 | 7 | 8 | 9 | 10 | 11 | 12 | 13 | 14 | 15 | 16 | 17 | 18 | 19 | 20 | 21 | 22 | 23 | 24 | 25 | 26 |
|---|---|---|---|---|---|---|---|---|---|---|---|---|---|---|---|---|---|---|---|---|---|---|---|---|---|---|---|
| **Processes** | Reversible | R | R | R | | | | | | | | | | | | | | | | | | | | | | | |
| | Irreversible | | | | | | | | | | | | | | R | R | R | R | | | R | R | | | | | |
| | Cyclic | | | | | R | R | R | R | | | | | | | | | | | | | | | | R | R | R |
| | Isothermal | | | | R | R | | | | | | | | | | | | | | | | | | | | | |
| | Isobaric | | | | | | | | | | | | | | | | | | | | | | | | | | |
| | Isochoric | | | | | | | | | | | | | | | | | | | | | | R | R | | | |
| | Adiabatic | R | R | | | | | | | | | | | | | | | | | | | | | | R | | |
| **Systems** | Systems & Universe | | | | | | | | | | | | | | | | | | | | | | | | | | |
| | Isolated System | | | | | | | | | | | | | R | R | R | R | R | | | | R | | R | | | |
| | Ideal Gas | | | | R | R | | | | | | R | | | | | | | | | | | | | | | |
| **Quantities and Relations** | Path Independence of State Variables | | | | | R | | R | R | | | R | | | | | | | | | | | | | R | R | R |
| | Internal Energy | | R | | R | R | R | | | R | | R | R | R | R | R | R | | | | | | R | | | R | R |
| | Relation to T | | | | R | R | | | | | | R | | | | | | | | | | | | | | | |
| | Heat | R | R | | R | R | | | R | | R | R | | R | R | R | | R | R | R | | | | | | | R |
| | Work | | R | R | R | R | | R | | R | R | R | R | | R | R | R | | | | | | | | | | R |
| | Relation to p, V | | R | R | R | R | | R | | R | R | R | R | | | | | | | | | | | | | | R |
| | Entropy | R | | | R | | R | | | | | | | | R | R | R | R | | | R | | | R | | | |
| | Relation to Q, T | R | | | R | | | | | | | | | | R | R | R | | | | | | | | | | |
| **Representation** | PV Diagram | | | | | | R | | R | R | R | R | | | | | | | | | | | | | | | R |
| **First Law** | 1st Law | | R | | R | R | | | R | | R | R | R | | R | R | R | | R | | | | R | | | | R |
| **Second Law** | 2nd Law | | | | | | | | | | | | | | R | R | R | R | | R | R | | | | | | |
| | Engine Efficiency | | | | | | | | | | | | | | | | | | | | | | | | | | |

**FIGURE 1a.** Topics for items 1-26 (topics for items 27-78 are in Figs. 1b and 1c). Along with other physics faculty members who teach thermodynamics, we identified topics involved in each item used for one possible method of solving the problem (note that several other approaches can be used to solve different problems). An "R" indicates that a concept is required to solve the problem using the cited solution method or the information is provided in the question.

| | Item # | 27 | 28 | 29 | 30 | 31 | 32 | 33 | 34 | 35 | 36 | 37 | 38 | 39 | 40 | 41 | 42 | 43 | 44 | 45 | 46 | 47 | 48 | 49 | 50 | 51 | 52 | 53 |
|---|---|---|---|---|---|---|---|---|---|---|---|---|---|---|---|---|---|---|---|---|---|---|---|---|---|---|---|---|
| **Processes** | Reversible | | R | | | | | | | R | R | | | | | | | | | | | | | | | | | |
| | Irreversible | R | | | | | | | | | | | | | | | | | | | | | | | | R | R | R |
| | Cyclic | | | R | R | | | | | | R | | | | | | | | | | | | | | | | | |
| | Isothermal | | | | | | R | | | | | | R | | | | | | | | | | | | | | | |
| | Isobaric | | | | | | | | | | | R | | | | R | | | | R | | | | | | | | |
| | Isochoric | | | | | | | | | | | | | | R | | R | | R | | | | | R | | | | R |
| | Adiabatic | | | | R | R | R | | | | | | | R | | | | | | | | | | | | | | R |
| **Systems** | Systems & Universe | | | | | | | | | R | R | R | R | | | | | | | | | | | | | | | |
| | Isolated System | R | | | | | | | | | | | | | | | | | | | | | | | | R | R | R | R | R |
| | Ideal Gas | | | | | | | R | R | | | R | R | | | | | | R | R | | | | | | | | |
| **Quantities and Relations** | Path Indep of State Variables | | | | | | | | | | R | | | | R | R | | | | | R | R | R | | | | | |
| | Internal Energy | | | | | | | R | R | | | | | | R | R | | | R | R | R | | | R | | | | |
| |    Relation to T | | | | | | | R | R | | | | | | R | R | | | R | R | | | | | | | | |
| | Heat | | | R | R | R | R | | | | | R | R | R | R | R | | | | | | | | | | R | R | |
| | Work | | | | R | | | | | | | | | | R | R | R | R | | R | | | R | | | | | |
| |    Relation to p, V | | | | | | | | | | | | | | R | R | R | R | | | | | | | | | | |
| | Entropy | R | | | | | | | | R | | | | | | | | | | | | | | R | | R | R | R |
| |    Relation to Q, T | | | | | | | | | R | | | | | | | | | | | | | | | | | R | R |
| **Representation** | PV Diagram | | | | | | | | | | | | | | R | R | | R | R | R | | | | | | | | |
| **First Law** | 1st Law | | | | | | | | | | | | | | R | R | | | | | | | | | | | | |
| **Second Law** | 2nd Law | R | R | R | | | | | | R | R | | | | | | | | | | | | | | | R | R | R |
| | Engine Efficiency | | R | | | | | | | R | | | | | | | | | | | | | | | | | | |

**FIGURE 1b.** Topics for items 27-53.

|   | Item # | 54 | 55 | 56 | 57 | 58 | 59 | 60 | 61 | 62 | 63 | 64 | 65 | 66 | 67 | 68 | 69 | 70 | 71 | 72 | 73 | 74 | 75 | 76 | 77 | 78 |
|---|---|---|---|---|---|---|---|---|---|---|---|---|---|---|---|---|---|---|---|---|---|---|---|---|---|---|
| **Processes** | Reversible |  |  |  |  |  |  |  |  |  |  | R |  |  |  |  |  |  |  |  |  |  |  |  |  |  |
|  | Irreversible |  |  | R |  |  |  |  |  |  |  |  | R |  |  |  |  |  |  |  | R | R | R |  |  | R |
|  | Cyclic |  |  |  |  |  |  |  |  |  |  |  |  |  |  |  |  |  |  |  |  |  |  |  |  |  |
|  | Isothermal |  |  |  |  |  |  | R | R | R |  | R | R |  |  |  |  |  |  |  | R | R |  |  |  |  |
|  | Isobaric |  |  |  |  |  |  |  |  |  |  |  |  |  |  |  |  |  |  |  |  |  |  |  |  |  |
|  | Isochoric |  |  |  |  |  |  |  |  |  |  |  |  |  | R |  | R | R | R |  |  |  |  |  |  |  |
|  | Adiabatic |  |  |  |  |  |  |  |  |  |  |  |  |  |  |  | R |  | R | R | R |  |  |  |  |  |
| **Systems** | Systems & Universe |  |  |  |  |  |  |  |  |  |  | R |  |  | R |  |  |  |  |  |  |  |  |  |  |  |
|  | Isolated System |  |  |  |  |  |  |  |  |  |  |  |  |  |  |  |  |  | R | R | R | R | R |  |  | R |
|  | Ideal Gas |  |  |  |  |  |  | R | R |  |  |  | R |  |  |  | R | R |  |  | R | R |  |  |  |  |
| **Quantities and Relations** | Path Indep of St Var |  |  |  |  |  |  |  |  |  |  |  |  |  |  |  |  |  |  |  |  |  |  |  |  |  |
|  | Internal Energy |  |  |  |  |  |  | R | R |  |  | R |  |  | R | R | R |  | R |  |  |  |  |  |  |  |
|  |    Relation to T |  |  |  |  |  |  | R | R |  |  |  | R |  |  | R | R |  |  |  |  |  |  |  |  |  |
|  | Heat | R | R | R |  |  |  | R |  | R |  | R |  | R |  | R |  | R |  |  |  |  |  | R | R | R |
|  | Work |  |  |  | R | R | R | R |  | R |  |  |  |  |  | R |  | R | R |  |  |  |  |  |  |  |
|  |    Relation to p, V |  |  |  | R | R | R | R |  | R |  |  |  |  |  | R |  |  |  |  |  |  |  |  |  |  |
|  | Entropy |  |  |  |  |  |  |  |  |  | R | R |  | R | R |  |  |  |  |  | R | R | R |  |  |  |
|  |    Relation to Q, T |  |  |  |  |  |  |  |  |  | R |  |  | R |  |  |  |  |  |  |  |  |  |  |  |  |
| **Representation** | PV Diagram |  |  |  | R | R | R |  |  |  |  |  |  |  |  |  |  |  |  |  |  |  |  |  |  |  |
| **First Law** | 1st Law |  | R |  |  |  | R |  |  |  |  |  |  |  |  | R |  | R |  |  |  |  |  |  | R |  |
| **Second Law** | 2nd Law |  |  | R |  |  |  |  |  |  |  | R |  | R |  |  |  |  |  |  | R | R | R |  |  | R |
|  | Engine Efficiency |  |  |  |  |  |  |  |  |  |  |  |  |  |  |  |  |  |  |  |  |  |  |  |  |  |

**FIGURE 1c.** Topics for items 54-78.

# APPENDIX B: STUDENT PERFORMANCE ON EACH QUESTION FOR IN-PERSON ADMINISTRATION

Appendix B shows the distribution of answer choices for each group on each question when the survey was given in class.

TABLE III: Average percentage scores for each of the four choices for each item for in-person administration of the STPFaSL-Long instrument for each group before (pre) and after (post) instruction in relevant concepts. Abbreviations for various student groups: Upper (students in junior/senior level thermodynamics and physics Ph.D. students in their first semester of a Ph.D. program who had also only taken the junior/senior level thermodynamics course), calc (students in introductory calculus-based physics courses), algebra (students in introductory algebra-based physics courses). The four columns after the item number show the percentage of students who selected choices A-D for each item. Questions that are true/false are coded with A being true and B being false; options C and D are not used. The number of students in each group is the same as in Table II. The correct answers are bold, underlined, and italicized.

| Item # | A | B | C | D | Level |
|---|---|---|---|---|---|
| 1 | 26% | 2% | ***69%*** | 3% | Upper Post |
|   | 37% | 11% | ***47%*** | 6% | Calc Post |
|   | 56% | 17% | ***24%*** | 3% | Calc Pre |
|   | 55% | 9% | ***31%*** | 5% | Algebra Post |
|   | 57% | 14% | ***25%*** | 4% | Algebra Pre |
| 2 | ***54%*** | 21% | 20% | 6% | Upper Post |
|   | ***30%*** | 28% | 39% | 4% | Calc Post |
|   | ***20%*** | 32% | 46% | 3% | Calc Pre |
|   | ***24%*** | 26% | 47% | 2% | Algebra Post |
|   | ***28%*** | 28% | 43% | 1% | Algebra Pre |
| 3 | ***75%*** | 10% | 11% | 3% | Upper Post |
|   | ***54%*** | 24% | 19% | 4% | Calc Post |
|   | ***48%*** | 19% | 27% | 6% | Calc Pre |
|   | ***41%*** | 27% | 27% | 5% | Algebra Post |
|   | ***41%*** | 25% | 28% | 6% | Algebra Pre |
| 4 | 13% | ***68%*** | 17% | 2% | Upper Post |
|   | 18% | ***30%*** | 50% | 3% | Calc Post |
|   | 27% | ***42%*** | 30% | 1% | Calc Pre |
|   | 18% | ***21%*** | 59% | 2% | Algebra Post |
|   | 19% | ***28%*** | 51% | 2% | Algebra Pre |
| 5 | 26% | ***42%*** | 31% | 1% | Upper Post |
|   | 29% | ***36%*** | 33% | 3% | Calc Post |
|   | 28% | ***50%*** | 20% | 3% | Calc Pre |
|   | 27% | ***48%*** | 22% | 3% | Algebra Post |
|   | 23% | ***54%*** | 22% | 2% | Algebra Pre |
| 6 | ***89%*** | 8% | 2% | 1% | Upper Post |
|   | ***80%*** | 9% | 9% | 2% | Calc Post |
|   | ***79%*** | 5% | 10% | 6% | Calc Pre |
|   | ***80%*** | 11% | 6% | 3% | Algebra Post |
|   | ***80%*** | 10% | 5% | 5% | Algebra Pre |
| 7 | 31% | _**51%**_ | 16% | 2% | Upper Post |
| | 31% | _**46%**_ | 21% | 2% | Calc Post |
| | 26% | _**13%**_ | 59% | 3% | Calc Pre |
| | 28% | _**18%**_ | 51% | 3% | Algebra Post |
| | 22% | _**5%**_ | 71% | 2% | Algebra Pre |
| 8 | 36% | 8% | _**49%**_ | 7% | Upper Post |
| | 25% | 16% | _**54%**_ | 5% | Calc Post |
| | 16% | 12% | _**67%**_ | 5% | Calc Pre |
| | 19% | 11% | _**65%**_ | 6% | Algebra Post |
| | 14% | 8% | _**75%**_ | 4% | Algebra Pre |
| 9 | 16% | 29% | _**44%**_ | 10% | Upper Post |
| | 33% | 26% | _**35%**_ | 6% | Calc Post |
| | 44% | 22% | _**21%**_ | 13% | Calc Pre |
| | 56% | 12% | _**20%**_ | 12% | Algebra Post |
| | 58% | 16% | _**16%**_ | 10% | Algebra Pre |
| 10 | 9% | _**88%**_ | 3% | 0% | Upper Post |
| | 18% | _**72%**_ | 9% | 1% | Calc Post |
| | 30% | _**55%**_ | 13% | 2% | Calc Pre |
| | 30% | _**58%**_ | 10% | 2% | Algebra Post |
| | 34% | _**54%**_ | 11% | 2% | Algebra Pre |
| 11 | 21% | _**66%**_ | 2% | 10% | Upper Post |
| | 23% | _**64%**_ | 9% | 3% | Calc Post |
| | 25% | _**56%**_ | 15% | 4% | Calc Pre |
| | 27% | _**55%**_ | 15% | 3% | Algebra Post |
| | 27% | _**58%**_ | 11% | 3% | Algebra Pre |
| 12 | 28% | 17% | _**47%**_ | 8% | Upper Post |
| | 56% | 11% | _**29%**_ | 4% | Calc Post |
| | 67% | 6% | _**23%**_ | 4% | Calc Pre |
| | 66% | 6% | _**23%**_ | 5% | Algebra Post |
| | 72% | 6% | _**17%**_ | 5% | Algebra Pre |
| 13 | 3% | 2% | _**94%**_ | 0% | Upper Post |
| | 6% | 9% | _**82%**_ | 3% | Calc Post |
| | 7% | 6% | _**84%**_ | 2% | Calc Pre |
| | 8% | 8% | _**83%**_ | 2% | Algebra Post |
| | 12% | 6% | _**79%**_ | 2% | Algebra Pre |
| 14 | 1% | _**93%**_ | 3% | 2% | Upper Post |
| | 5% | _**91%**_ | 3% | 1% | Calc Post |
| | 8% | _**84%**_ | 6% | 2% | Calc Pre |
| | 5% | _**90%**_ | 5% | 1% | Algebra Post |
| | 5% | _**88%**_ | 6% | 1% | Algebra Pre |
| 15 | **_97%_** | 3% | 0% | 0% | Upper Post |
| | **_85%_** | 9% | 6% | 1% | Calc Post |
| | **_77%_** | 11% | 9% | 3% | Calc Pre |
| | **_86%_** | 8% | 5% | 1% | Algebra Post |
| | **_80%_** | 9% | 10% | 1% | Algebra Pre |
| 16 | 17% | **_74%_** | 6% | 3% | Upper Post |
| | 17% | **_71%_** | 9% | 3% | Calc Post |
| | 16% | **_69%_** | 13% | 3% | Calc Pre |
| | 11% | **_80%_** | 8% | 1% | Algebra Post |
| | 13% | **_74%_** | 12% | 1% | Algebra Pre |
| 17 | **_78%_** | 5% | 15% | 2% | Upper Post |
| | **_23%_** | 5% | 69% | 4% | Calc Post |
| | **_16%_** | 6% | 75% | 3% | Calc Pre |
| | **_20%_** | 7% | 71% | 1% | Algebra Post |
| | **_12%_** | 6% | 79% | 3% | Algebra Pre |
| 18 | 4% | **_96%_** | - | - | Upper Post |
| | 17% | **_83%_** | - | - | Calc Post |
| | 18% | **_82%_** | - | - | Calc Pre |
| | 21% | **_79%_** | - | - | Algebra Post |
| | 28% | **_72%_** | - | - | Algebra Pre |
| 19 | 19% | **_81%_** | - | - | Upper Post |
| | 53% | **_47%_** | - | - | Calc Post |
| | 69% | **_31%_** | - | - | Calc Pre |
| | 51% | **_49%_** | - | - | Algebra Post |
| | 68% | **_32%_** | - | - | Algebra Pre |
| 20 | **_82%_** | 18% | - | - | Upper Post |
| | **_77%_** | 23% | - | - | Calc Post |
| | **_73%_** | 27% | - | - | Calc Pre |
| | **_80%_** | 20% | - | - | Algebra Post |
| | **_68%_** | 32% | - | - | Algebra Pre |
| 21 | 8% | **_87%_** | 6% | 0% | Upper Post |
| | 29% | **_47%_** | 21% | 2% | Calc Post |
| | 30% | **_43%_** | 24% | 3% | Calc Pre |
| | 27% | **_50%_** | 22% | 0% | Algebra Post |
| | 33% | **_53%_** | 12% | 3% | Algebra Pre |
| 22 | **_63%_** | 34% | 3% | 0% | Upper Post |
| | **_37%_** | 45% | 17% | 1% | Calc Post |
| | **_39%_** | 37% | 20% | 4% | Calc Pre |
| | **_39%_** | 37% | 20% | 3% | Algebra Post |
| | **_36%_** | 39% | 23% | 2% | Algebra Pre |
| 23 | **_76%_** | 6% | 15% | 3% | Upper Post |
| | **_57%_** | 14% | 28% | 1% | Calc Post |
| | **_60%_** | 10% | 29% | 2% | Calc Pre |
| | **_58%_** | 11% | 27% | 4% | Algebra Post |
| | **_55%_** | 13% | 31% | 1% | Algebra Pre |
| 24 | **_57%_** | 38% | 0% | 4% | Upper Post |
| | **_63%_** | 27% | 8% | 2% | Calc Post |
| | **_63%_** | 19% | 13% | 6% | Calc Pre |
| | **_67%_** | 22% | 9% | 2% | Algebra Post |
| | **_77%_** | 16% | 5% | 2% | Algebra Pre |
| 25 | 6% | 9% | **_83%_** | 2% | Upper Post |
| | 17% | 11% | **_72%_** | 1% | Calc Post |
| | 15% | 19% | **_63%_** | 4% | Calc Pre |
| | 14% | 15% | **_69%_** | 2% | Algebra Post |
| | 13% | 13% | **_72%_** | 2% | Algebra Pre |
| 26 | 25% | **_52%_** | 13% | 10% | Upper Post |
| | 37% | **_31%_** | 28% | 4% | Calc Post |
| | 49% | **_16%_** | 24% | 11% | Calc Pre |
| | 55% | **_16%_** | 24% | 5% | Algebra Post |
| | 61% | **_18%_** | 17% | 3% | Algebra Pre |
| 27 | 27% | **_73%_** | - | - | Upper Post |
| | 71% | **_29%_** | - | - | Calc Post |
| | 67% | **_33%_** | - | - | Calc Pre |
| | 65% | **_35%_** | - | - | Algebra Post |
| | 70% | **_30%_** | - | - | Algebra Pre |
| 28 | **_83%_** | 17% | - | - | Upper Post |
| | **_65%_** | 35% | - | - | Calc Post |
| | **_56%_** | 44% | - | - | Calc Pre |
| | **_67%_** | 33% | - | - | Algebra Post |
| | **_46%_** | 54% | - | - | Algebra Pre |
| 29 | **_89%_** | 11% | - | - | Upper Post |
| | **_74%_** | 26% | - | - | Calc Post |
| | **_73%_** | 27% | - | - | Calc Pre |
| | **_77%_** | 23% | - | - | Algebra Post |
| | **_67%_** | 33% | - | - | Algebra Pre |
| 30 | 15% | **85%** | - | - | Upper Post |
| | 33% | **67%** | - | - | Calc Post |
| | 45% | **55%** | - | - | Calc Pre |
| | 45% | **55%** | - | - | Algebra Post |
| | 50% | **50%** | - | - | Algebra Pre |
| 31 | **_87%_** | 13% | - | - | Upper Post |
| | **_69%_** | 31% | - | - | Calc Post |
| | **_51%_** | 49% | - | - | Calc Pre |
| | **_69%_** | 31% | - | - | Algebra Post |
| | **_53%_** | 47% | - | - | Algebra Pre |
| 32 | 16% | **_84%_** | - | - | Upper Post |
| | 46% | **_54%_** | - | - | Calc Post |
| | 55% | **_45%_** | - | - | Calc Pre |
| | 50% | **_50%_** | - | - | Algebra Post |
| | 60% | **_40%_** | - | - | Algebra Pre |
| 33 | 8% | **_83%_** | 7% | 2% | Upper Post |
| | 16% | **_76%_** | 8% | 0% | Calc Post |
| | 9% | **_84%_** | 6% | 1% | Calc Pre |
| | 11% | **_81%_** | 6% | 1% | Algebra Post |
| | 8% | **_85%_** | 6% | 1% | Algebra Pre |
| 34 | 7% | 15% | **_70%_** | 9% | Upper Post |
| | 12% | 25% | **_58%_** | 4% | Calc Post |
| | 24% | 15% | **_59%_** | 2% | Calc Pre |
| | 20% | 20% | **_55%_** | 5% | Algebra Post |
| | 24% | 17% | **_58%_** | 1% | Algebra Pre |
| 35 | 16% | **_84%_** | - | - | Upper Post |
| | 30% | **_70%_** | - | - | Calc Post |
| | 19% | **_81%_** | - | - | Calc Pre |
| | 26% | **_74%_** | - | - | Algebra Post |
| | 22% | **_78%_** | - | - | Algebra Pre |
| 36 | 26% | 21% | 32% | **_21%_** | Upper Post |
| | 19% | 11% | 30% | **_40%_** | Calc Post |
| | 19% | 13% | 35% | **_34%_** | Calc Pre |
| | 21% | 15% | 29% | **_35%_** | Algebra Post |
| | 17% | 13% | 34% | **_37%_** | Algebra Pre |
| 37 | **_82%_** | 18% | - | - | Upper Post |
| | **_55%_** | 45% | - | - | Calc Post |
| | **_55%_** | 45% | - | - | Calc Pre |
| | **_45%_** | 55% | - | - | Algebra Post |
| | **_48%_** | 52% | - | - | Algebra Pre |
| 38 | 22% | **_78%_** | - | - | Upper Post |
| | 28% | **_72%_** | - | - | Calc Post |
| | 53% | **_47%_** | - | - | Calc Pre |
| | 33% | **_67%_** | - | - | Algebra Post |
| | 45% | **_55%_** | - | - | Algebra Pre |
|    | 13% | **_88%_** | - | - | Upper Post |
|    | 39% | **_61%_** | - | - | Calc Post |
| 39 | 48% | **_52%_** | - | - | Calc Pre |
|    | 52% | **_48%_** | - | - | Algebra Post |
|    | 56% | **_44%_** | - | - | Algebra Pre |
|    | 9%  | **_83%_** | 3%  | 5% | Upper Post |
|    | 18% | **_61%_** | 18% | 4% | Calc Post |
| 40 | 11% | **_55%_** | 27% | 6% | Calc Pre |
|    | 13% | **_58%_** | 24% | 6% | Algebra Post |
|    | 14% | **_60%_** | 20% | 6% | Algebra Pre |
|    | 9%  | **_67%_** | 17% | 7% | Upper Post |
|    | 17% | **_45%_** | 32% | 6% | Calc Post |
| 41 | 25% | **_37%_** | 31% | 7% | Calc Pre |
|    | 20% | **_42%_** | 31% | 7% | Algebra Post |
|    | 24% | **_43%_** | 26% | 7% | Algebra Pre |
|    | 9%  | 7%  | **_84%_** | 0% | Upper Post |
|    | 15% | 10% | **_74%_** | 1% | Calc Post |
| 42 | 50% | 18% | **_28%_** | 3% | Calc Pre |
|    | 38% | 14% | **_46%_** | 2% | Algebra Post |
|    | 64% | 15% | **_17%_** | 4% | Algebra Pre |
|    | **_83%_** | 12% | 4%  | 0% | Upper Post |
|    | **_74%_** | 16% | 9%  | 1% | Calc Post |
| 43 | **_45%_** | 26% | 26% | 3% | Calc Pre |
|    | **_60%_** | 21% | 17% | 2% | Algebra Post |
|    | **_43%_** | 27% | 28% | 2% | Algebra Pre |
|    | 7%  | **_88%_** | 4%  | 1% | Upper Post |
|    | 18% | **_69%_** | 8%  | 4% | Calc Post |
| 44 | 15% | **_69%_** | 12% | 4% | Calc Pre |
|    | 17% | **_70%_** | 11% | 2% | Algebra Post |
|    | 14% | **_74%_** | 8%  | 3% | Algebra Pre |
|    | 7%  | **_65%_** | 20% | 8% | Upper Post |
|    | 17% | **_43%_** | 33% | 7% | Calc Post |
| 45 | 29% | **_32%_** | 34% | 5% | Calc Pre |
|    | 22% | **_37%_** | 37% | 4% | Algebra Post |
|    | 29% | **_32%_** | 36% | 3% | Algebra Pre |
|    | 10% | **_90%_** | - | - | Upper Post |
|    | 26% | **_74%_** | - | - | Calc Post |
| 46 | 37% | **_63%_** | - | - | Calc Pre |
|    | 32% | **_68%_** | - | - | Algebra Post |
|    | 35% | **_65%_** | - | - | Algebra Pre |
| 47 | **_84%_** | 16% | - | - | Upper Post |
| | **_71%_** | 29% | - | - | Calc Post |
| | **_72%_** | 28% | - | - | Calc Pre |
| | **_72%_** | 28% | - | - | Algebra Post |
| | **_73%_** | 27% | - | - | Algebra Pre |
| 48 | **_64%_** | 36% | - | - | Upper Post |
| | **_69%_** | 31% | - | - | Calc Post |
| | **_66%_** | 34% | - | - | Calc Pre |
| | **_71%_** | 29% | - | - | Algebra Post |
| | **_72%_** | 28% | - | - | Algebra Pre |
| 49 | **_89%_** | 9% | 2% | 0% | Upper Post |
| | **_72%_** | 17% | 6% | 6% | Calc Post |
| | **_79%_** | 12% | 6% | 3% | Calc Pre |
| | **_75%_** | 16% | 7% | 3% | Algebra Post |
| | **_78%_** | 13% | 3% | 6% | Algebra Pre |
| 50 | 8% | 3% | **_89%_** | 0% | Upper Post |
| | 17% | 13% | **_69%_** | 2% | Calc Post |
| | 7% | 12% | **_78%_** | 3% | Calc Pre |
| | 7% | 14% | **_77%_** | 2% | Algebra Post |
| | 7% | 11% | **_78%_** | 5% | Algebra Pre |
| 51 | **_93%_** | 3% | 3% | 0% | Upper Post |
| | **_78%_** | 11% | 7% | 3% | Calc Post |
| | **_72%_** | 14% | 12% | 2% | Calc Pre |
| | **_81%_** | 12% | 7% | 1% | Algebra Post |
| | **_73%_** | 14% | 11% | 2% | Algebra Pre |
| 52 | 12% | **_80%_** | 6% | 2% | Upper Post |
| | 16% | **_78%_** | 6% | 1% | Calc Post |
| | 17% | **_70%_** | 10% | 4% | Calc Pre |
| | 14% | **_79%_** | 6% | 1% | Algebra Post |
| | 14% | **_77%_** | 6% | 3% | Algebra Pre |
| 53 | **_84%_** | 3% | 11% | 1% | Upper Post |
| | **_23%_** | 10% | 62% | 5% | Calc Post |
| | **_14%_** | 8% | 74% | 4% | Calc Pre |
| | **_21%_** | 10% | 68% | 1% | Algebra Post |
| | **_14%_** | 13% | 68% | 5% | Algebra Pre |
| 54 | 8% | **_92%_** | - | - | Upper Post |
| | 25% | **_75%_** | - | - | Calc Post |
| | 24% | **_76%_** | - | - | Calc Pre |
| | 23% | **_77%_** | - | - | Algebra Post |
| | 30% | **_70%_** | - | - | Algebra Pre |
| 55 | 22% | _**78%**_ | - | - | Upper Post |
| | 52% | _**48%**_ | - | - | Calc Post |
| | 63% | _**37%**_ | - | - | Calc Pre |
| | 46% | _**54%**_ | - | - | Algebra Post |
| | 57% | _**43%**_ | - | - | Algebra Pre |
| 56 | _**84%**_ | 16% | - | - | Upper Post |
| | _**69%**_ | 31% | - | - | Calc Post |
| | _**70%**_ | 30% | - | - | Calc Pre |
| | _**79%**_ | 21% | - | - | Algebra Post |
| | _**71%**_ | 29% | - | - | Algebra Pre |
| 57 | 3% | _**89%**_ | 7% | 1% | Upper Post |
| | 10% | _**64%**_ | 24% | 1% | Calc Post |
| | 24% | _**28%**_ | 43% | 5% | Calc Pre |
| | 15% | _**52%**_ | 33% | 1% | Algebra Post |
| | 21% | _**27%**_ | 49% | 4% | Algebra Pre |
| 58 | 6% | _**87%**_ | 7% | 1% | Upper Post |
| | 10% | _**61%**_ | 27% | 2% | Calc Post |
| | 10% | _**32%**_ | 54% | 5% | Calc Pre |
| | 13% | _**44%**_ | 41% | 2% | Algebra Post |
| | 17% | _**23%**_ | 59% | 2% | Algebra Pre |
| 59 | 9% | _**78%**_ | 12% | 1% | Upper Post |
| | 17% | _**54%**_ | 26% | 3% | Calc Post |
| | 14% | _**29%**_ | 51% | 6% | Calc Pre |
| | 14% | _**38%**_ | 44% | 3% | Algebra Post |
| | 11% | _**30%**_ | 53% | 7% | Algebra Pre |
| 60 | 12% | _**75%**_ | 10% | 2% | Upper Post |
| | 30% | _**43%**_ | 20% | 7% | Calc Post |
| | 39% | _**34%**_ | 24% | 3% | Calc Pre |
| | 37% | _**44%**_ | 15% | 4% | Algebra Post |
| | 37% | _**35%**_ | 24% | 4% | Algebra Pre |
| 61 | _**64%**_ | 24% | 7% | 6% | Upper Post |
| | _**48%**_ | 26% | 22% | 4% | Calc Post |
| | _**34%**_ | 36% | 26% | 4% | Calc Pre |
| | _**28%**_ | 39% | 28% | 5% | Algebra Post |
| | _**27%**_ | 44% | 26% | 3% | Algebra Pre |
| 62 | _**81%**_ | 11% | 4% | 3% | Upper Post |
| | _**65%**_ | 19% | 11% | 4% | Calc Post |
| | _**46%**_ | 28% | 22% | 5% | Calc Pre |
| | _**55%**_ | 24% | 16% | 5% | Algebra Post |
| | _**47%**_ | 26% | 21% | 6% | Algebra Pre |
| 63 | **_61%_** | 4% | 35% | 0% | Upper Post |
| | **_39%_** | 20% | 38% | 3% | Calc Post |
| | **_50%_** | 23% | 25% | 3% | Calc Pre |
| | **_54%_** | 22% | 23% | 1% | Algebra Post |
| | **_55%_** | 19% | 25% | 1% | Algebra Pre |
| 64 | **_52%_** | 45% | 3% | 0% | Upper Post |
| | **_49%_** | 28% | 19% | 4% | Calc Post |
| | **_48%_** | 27% | 20% | 4% | Calc Pre |
| | **_56%_** | 26% | 15% | 4% | Algebra Post |
| | **_50%_** | 27% | 17% | 7% | Algebra Pre |
| 65 | **_69%_** | 22% | 6% | 3% | Upper Post |
| | **_49%_** | 27% | 23% | 2% | Calc Post |
| | **_34%_** | 32% | 30% | 4% | Calc Pre |
| | **_40%_** | 35% | 21% | 4% | Algebra Post |
| | **_36%_** | 38% | 25% | 1% | Algebra Pre |
| 66 | **_88%_** | 7% | 2% | 3% | Upper Post |
| | **_55%_** | 22% | 20% | 3% | Calc Post |
| | **_58%_** | 22% | 17% | 4% | Calc Pre |
| | **_58%_** | 20% | 19% | 3% | Algebra Post |
| | **_59%_** | 20% | 18% | 3% | Algebra Pre |
| 67 | 14% | **_83%_** | 2% | 1% | Upper Post |
| | 38% | **_40%_** | 18% | 4% | Calc Post |
| | 44% | **_34%_** | 19% | 4% | Calc Pre |
| | 34% | **_47%_** | 17% | 2% | Algebra Post |
| | 38% | **_41%_** | 17% | 4% | Algebra Pre |
| 68 | 14% | **_70%_** | 9% | 7% | Upper Post |
| | 24% | **_55%_** | 17% | 4% | Calc Post |
| | 17% | **_55%_** | 23% | 5% | Calc Pre |
| | 18% | **_61%_** | 17% | 4% | Algebra Post |
| | 18% | **_58%_** | 19% | 6% | Algebra Pre |
| 69 | **_60%_** | 22% | 15% | 3% | Upper Post |
| | **_32%_** | 41% | 25% | 2% | Calc Post |
| | **_34%_** | 44% | 20% | 2% | Calc Pre |
| | **_25%_** | 51% | 20% | 4% | Algebra Post |
| | **_27%_** | 50% | 21% | 2% | Algebra Pre |
| 70 | **_29%_** | 45% | 20% | 7% | Upper Post |
| | **_29%_** | 48% | 19% | 4% | Calc Post |
| | **_28%_** | 42% | 25% | 6% | Calc Pre |
| | **_25%_** | 48% | 21% | 6% | Algebra Post |
| | **_30%_** | 46% | 21% | 4% | Algebra Pre |
| 71 | **_71%_** | 23% | 5% | 1% | Upper Post |
| | **_67%_** | 25% | 4% | 3% | Calc Post |
| | **_75%_** | 16% | 5% | 4% | Calc Pre |
| | **_70%_** | 21% | 7% | 2% | Algebra Post |
| | **_77%_** | 14% | 7% | 3% | Algebra Pre |
| 72 | 7% | 13% | **_78%_** | 2% | Upper Post |
| | 21% | 17% | **_60%_** | 2% | Calc Post |
| | 15% | 11% | **_72%_** | 2% | Calc Pre |
| | 17% | 17% | **_63%_** | 3% | Algebra Post |
| | 15% | 17% | **_66%_** | 3% | Algebra Pre |
| 73 | **_79%_** | 9% | 11% | 1% | Upper Post |
| | **_40%_** | 23% | 30% | 8% | Calc Post |
| | **_37%_** | 24% | 30% | 9% | Calc Pre |
| | **_48%_** | 23% | 24% | 5% | Algebra Post |
| | **_39%_** | 19% | 31% | 11% | Algebra Pre |
| 74 | **_85%_** | 7% | 6% | 2% | Upper Post |
| | **_40%_** | 24% | 26% | 9% | Calc Post |
| | **_38%_** | 25% | 29% | 8% | Calc Pre |
| | **_49%_** | 22% | 23% | 6% | Algebra Post |
| | **_35%_** | 25% | 30% | 11% | Algebra Pre |
| 75 | **_85%_** | 7% | 8% | 0% | Upper Post |
| | **_35%_** | 15% | 47% | 3% | Calc Post |
| | **_30%_** | 14% | 54% | 2% | Calc Pre |
| | **_43%_** | 13% | 43% | 1% | Algebra Post |
| | **_30%_** | 19% | 49% | 2% | Algebra Pre |
| 76 | 10% | **_90%_** | - | - | Upper Post |
| | 36% | **_64%_** | - | - | Calc Post |
| | 40% | **_60%_** | - | - | Calc Pre |
| | 34% | **_66%_** | - | - | Algebra Post |
| | 48% | **_52%_** | - | - | Algebra Pre |
| 77 | 21% | **_79%_** | - | - | Upper Post |
| | 56% | **_44%_** | - | - | Calc Post |
| | 55% | **_45%_** | - | - | Calc Pre |
| | 52% | **_48%_** | - | - | Algebra Post |
| | 59% | **_41%_** | - | - | Algebra Pre |
| 78 | **_87%_** | 13% | - | - | Upper Post |
| | **_68%_** | 32% | - | - | Calc Post |
| | **_70%_** | 30% | - | - | Calc Pre |
| | **_67%_** | 33% | - | - | Algebra Post |
| | **_65%_** | 35% | - | - | Algebra Pre |

**APPENDIX C: STUDENT PERFORMANCE ON EACH QUESTION FOR ONLINE ADMINISTRATION**

Appendix C shows the distribution of answer choices for each group on each question when the survey was given online.

TABLE IV. Average percentage scores for each of the four choices for each item for online administration of the STPFaSL-Long instrument for each group after instruction in relevant concepts. Abbreviations for various student groups: Upper (students in junior/senior level thermodynamics and physics Ph.D. students in their first semester of a Ph.D. program who had also only taken the junior/senior level thermodynamics course), Calc (students in introductory calculus-based physics courses), Algebra (students in introductory algebra-based physics courses). The four columns after the item number show the percentage of students who selected choices A-D for each item. The number of students in each group is the same as in Table II.

| Item # | A | B | C | D | Level |
|---|---|---|---|---|---|
| 1 | 18% | 5% | **_68%_** | 9% | Upper Post |
|   | 25% | 3% | **_64%_** | 8% | Calc Post |
|   | 33% | 7% | **_55%_** | 5% | Algebra Post |
| 2 | **_50%_** | 20% | 27% | 3% | Upper Post |
|   | **_61%_** | 10% | 27% | 2% | Calc Post |
|   | **_46%_** | 20% | 31% | 3% | Algebra Post |
| 3 | **_52%_** | 31% | 12% | 5% | Upper Post |
|   | **_65%_** | 20% | 13% | 3% | Calc Post |
|   | **_51%_** | 26% | 19% | 4% | Algebra Post |
| 4 | 12% | **_58%_** | 27% | 3% | Upper Post |
|   | 16% | **_57%_** | 24% | 2% | Calc Post |
|   | 21% | **_44%_** | 33% | 2% | Algebra Post |
| 5 | 20% | **_34%_** | 43% | 3% | Upper Post |
|   | 22% | **_57%_** | 20% | 2% | Calc Post |
|   | 22% | **_50%_** | 26% | 1% | Algebra Post |
| 6 | **_81%_** | 11% | 3% | 4% | Upper Post |
|   | **_86%_** | 7% | 5% | 2% | Calc Post |
|   | **_72%_** | 14% | 11% | 3% | Algebra Post |
| 7 | 30% | **_45%_** | 22% | 3% | Upper Post |
|   | 30% | **_50%_** | 16% | 4% | Calc Post |
|   | 24% | **_43%_** | 31% | 2% | Algebra Post |
| 8 | 40% | 9% | **_45%_** | 6% | Upper Post |
|   | 19% | 7% | **_71%_** | 2% | Calc Post |
|   | 24% | 12% | **_59%_** | 5% | Algebra Post |
| 9 | 20% | 26% | **_39%_** | 14% | Upper Post |
|   | 25% | 16% | **_51%_** | 8% | Calc Post |
|   | 31% | 23% | **_40%_** | 6% | Algebra Post |
| 10 | 18% | **_70%_** | 13% | 0% | Upper Post |
|   | 11% | **_82%_** | 6% | 1% | Calc Post |
|   | 20% | **_70%_** | 8% | 2% | Algebra Post |
| 11 | 23% | **_62%_** | 8% | 7% | Upper Post |
| | 12% | **_77%_** | 7% | 4% | Calc Post |
| | 21% | **_63%_** | 12% | 4% | Algebra Post |
| 12 | 33% | 22% | **_35%_** | 10% | Upper Post |
| | 39% | 9% | **_48%_** | 4% | Calc Post |
| | 41% | 15% | **_41%_** | 2% | Algebra Post |
| 13 | 12% | 6% | **_78%_** | 4% | Upper Post |
| | 11% | 13% | **_75%_** | 1% | Calc Post |
| | 13% | 17% | **_69%_** | 2% | Algebra Post |
| 14 | 6% | **_87%_** | 7% | 0% | Upper Post |
| | 5% | **_82%_** | 12% | 1% | Calc Post |
| | 10% | **_75%_** | 14% | 1% | Algebra Post |
| 15 | **_87%_** | 7% | 5% | 1% | Upper Post |
| | **_83%_** | 9% | 6% | 1% | Calc Post |
| | **_76%_** | 13% | 10% | 2% | Algebra Post |
| 16 | 19% | **_72%_** | 7% | 1% | Upper Post |
| | 15% | **_71%_** | 13% | 1% | Calc Post |
| | 16% | **_69%_** | 12% | 2% | Algebra Post |
| 17 | **_78%_** | 2% | 17% | 3% | Upper Post |
| | **_52%_** | 7% | 38% | 3% | Calc Post |
| | **_37%_** | 9% | 51% | 2% | Algebra Post |
| 18 | 15% | **_85%_** | - | - | Upper Post |
| | 14% | **_86%_** | - | - | Calc Post |
| | 25% | **_75%_** | - | - | Algebra Post |
| 19 | 32% | **_68%_** | - | - | Upper Post |
| | 37% | **_63%_** | - | - | Calc Post |
| | 32% | **_68%_** | - | - | Algebra Post |
| 20 | **_85%_** | 15% | - | - | Upper Post |
| | **_75%_** | 25% | - | - | Calc Post |
| | **_76%_** | 24% | - | - | Algebra Post |
| 21 | 10% | **_86%_** | 4% | 1% | Upper Post |
| | 12% | **_78%_** | 10% | 1% | Calc Post |
| | 20% | **_70%_** | 10% | 1% | Algebra Post |
| 22 | **_51%_** | 28% | 19% | 2% | Upper Post |
| | **_57%_** | 32% | 8% | 3% | Calc Post |
| | **_45%_** | 37% | 16% | 2% | Algebra Post |
| 23 | **_61%_** | 10% | 25% | 3% | Upper Post |
| | **_67%_** | 14% | 17% | 2% | Calc Post |
| | **_53%_** | 17% | 28% | 2% | Algebra Post |
| 24 | **_57%_** | 38% | 3% | 2% | Upper Post |
|  | **_75%_** | 17% | 6% | 2% | Calc Post |
|  | **_68%_** | 20% | 10% | 1% | Algebra Post |
| 25 | 13% | 6% | **_81%_** | 1% | Upper Post |
|  | 9% | 7% | **_83%_** | 1% | Calc Post |
|  | 15% | 17% | **_66%_** | 1% | Algebra Post |
| 26 | 31% | **_36%_** | 27% | 6% | Upper Post |
|  | 35% | **_44%_** | 19% | 3% | Calc Post |
|  | 36% | **_42%_** | 20% | 2% | Algebra Post |
| 27 | 44% | **_56%_** | - | - | Upper Post |
|  | 50% | **_50%_** | - | - | Calc Post |
|  | 60% | **_40%_** | - | - | Algebra Post |
| 28 | **_82%_** | 18% | - | - | Upper Post |
|  | **_79%_** | 21% | - | - | Calc Post |
|  | **_77%_** | 23% | - | - | Algebra Post |
| 29 | **_85%_** | 15% | - | - | Upper Post |
|  | **_89%_** | 11% | - | - | Calc Post |
|  | **_76%_** | 24% | - | - | Algebra Post |
| 30 | 26% | **_74%_** | - | - | Upper Post |
|  | 34% | **_66%_** | - | - | Calc Post |
|  | 38% | **_62%_** | - | - | Algebra Post |
| 31 | **_82%_** | 18% | - | - | Upper Post |
|  | **_89%_** | 11% | - | - | Calc Post |
|  | **_82%_** | 18% | - | - | Algebra Post |
| 32 | 23% | **_77%_** | - | - | Upper Post |
|  | 20% | **_80%_** | - | - | Calc Post |
|  | 30% | **_70%_** | - | - | Algebra Post |
| 33 | 8% | **_88%_** | 4% | 1% | Upper Post |
|  | 8% | **_85%_** | 6% | 1% | Calc Post |
|  | 12% | **_78%_** | 10% | 1% | Algebra Post |
| 34 | 11% | 22% | **_61%_** | 6% | Upper Post |
|  | 8% | 22% | **_65%_** | 5% | Calc Post |
|  | 16% | 26% | **_56%_** | 3% | Algebra Post |
| 35 | 25% | **_75%_** | - | - | Upper Post |
|  | 20% | **_80%_** | - | - | Calc Post |
|  | 32% | **_68%_** | - | - | Algebra Post |
| 36 | 24% | 22% | 24% | **_30%_** | Upper Post |
|  | 17% | 12% | 23% | **_47%_** | Calc Post |
|  | 14% | 19% | 23% | **_44%_** | Algebra Post |

| # | | | | | |
|---|---|---|---|---|---|
| 37 | **_69%_** | 31% | - | - | Upper Post |
|  | **_71%_** | 29% | - | - | Calc Post |
|  | **_60%_** | 40% | - | - | Algebra Post |
| 38 | 25% | **_75%_** | - | - | Upper Post |
|  | 17% | **_83%_** | - | - | Calc Post |
|  | 28% | **_72%_** | - | - | Algebra Post |
| 39 | 32% | **_68%_** | - | - | Upper Post |
|  | 31% | **_69%_** | - | - | Calc Post |
|  | 42% | **_58%_** | - | - | Algebra Post |
| 40 | 11% | **_72%_** | 13% | 4% | Upper Post |
|  | 7% | **_73%_** | 15% | 4% | Calc Post |
|  | 16% | **_60%_** | 18% | 5% | Algebra Post |
| 41 | 17% | **_55%_** | 20% | 8% | Upper Post |
|  | 17% | **_48%_** | 28% | 7% | Calc Post |
|  | 18% | **_45%_** | 32% | 6% | Algebra Post |
| 42 | 13% | 9% | **_77%_** | 1% | Upper Post |
|  | 12% | 9% | **_77%_** | 2% | Calc Post |
|  | 21% | 21% | **_57%_** | 2% | Algebra Post |
| 43 | **_56%_** | 40% | 4% | 1% | Upper Post |
|  | **_77%_** | 14% | 8% | 2% | Calc Post |
|  | **_60%_** | 24% | 15% | 1% | Algebra Post |
| 44 | 7% | **_79%_** | 11% | 3% | Upper Post |
|  | 11% | **_74%_** | 10% | 5% | Calc Post |
|  | 20% | **_58%_** | 19% | 4% | Algebra Post |
| 45 | 7% | **_58%_** | 30% | 5% | Upper Post |
|  | 13% | **_44%_** | 32% | 11% | Calc Post |
|  | 21% | **_38%_** | 36% | 6% | Algebra Post |
| 46 | 15% | **_85%_** | - | - | Upper Post |
|  | 18% | **_82%_** | - | - | Calc Post |
|  | 37% | **_63%_** | - | - | Algebra Post |
| 47 | **_83%_** | 17% | - | - | Upper Post |
|  | **_85%_** | 15% | - | - | Calc Post |
|  | **_73%_** | 27% | - | - | Algebra Post |
| 48 | **_75%_** | 25% | - | - | Upper Post |
|  | **_86%_** | 14% | - | - | Calc Post |
|  | **_77%_** | 23% | - | - | Algebra Post |
| 49 | **_78%_** | 13% | 5% | 3% | Upper Post |
|  | **_74%_** | 18% | 7% | 1% | Calc Post |
|  | **_62%_** | 25% | 11% | 2% | Algebra Post |

| # | | | | | |
|---|---|---|---|---|---|
| 50 | 16% | 8% | **_73%_** | 3% | Upper Post |
| | 12% | 13% | **_74%_** | 1% | Calc Post |
| | 15% | 22% | **_61%_** | 2% | Algebra Post |
| 51 | **_88%_** | 8% | 4% | 1% | Upper Post |
| | **_83%_** | 9% | 7% | 1% | Calc Post |
| | **_72%_** | 15% | 12% | 1% | Algebra Post |
| 52 | 16% | **_78%_** | 6% | 1% | Upper Post |
| | 11% | **_73%_** | 15% | 1% | Calc Post |
| | 17% | **_69%_** | 12% | 2% | Algebra Post |
| 53 | **_75%_** | 7% | 16% | 2% | Upper Post |
| | **_51%_** | 9% | 38% | 2% | Calc Post |
| | **_38%_** | 14% | 45% | 2% | Algebra Post |
| 54 | 16% | **_84%_** | - | - | Upper Post |
| | 19% | **_81%_** | - | - | Calc Post |
| | 26% | **_74%_** | - | - | Algebra Post |
| 55 | 31% | **_69%_** | - | - | Upper Post |
| | 35% | **_65%_** | - | - | Calc Post |
| | 33% | **_67%_** | - | - | Algebra Post |
| 56 | **_85%_** | 15% | - | - | Upper Post |
| | **_82%_** | 18% | - | - | Calc Post |
| | **_77%_** | 23% | - | - | Algebra Post |
| 57 | 7% | **_69%_** | 24% | 0% | Upper Post |
| | 5% | **_75%_** | 19% | 1% | Calc Post |
| | 12% | **_64%_** | 23% | 1% | Algebra Post |
| 58 | 5% | **_65%_** | 29% | 1% | Upper Post |
| | 6% | **_66%_** | 25% | 2% | Calc Post |
| | 10% | **_50%_** | 37% | 2% | Algebra Post |
| 59 | 16% | **_62%_** | 21% | 1% | Upper Post |
| | 7% | **_78%_** | 12% | 2% | Calc Post |
| | 18% | **_62%_** | 18% | 2% | Algebra Post |
| 60 | 18% | **_61%_** | 21% | 1% | Upper Post |
| | 15% | **_61%_** | 18% | 5% | Calc Post |
| | 25% | **_50%_** | 23% | 2% | Algebra Post |
| 61 | **_55%_** | 26% | 18% | 1% | Upper Post |
| | **_69%_** | 15% | 14% | 2% | Calc Post |
| | **_50%_** | 25% | 22% | 3% | Algebra Post |
| 62 | **_62%_** | 30% | 7% | 2% | Upper Post |
| | **_80%_** | 10% | 6% | 4% | Calc Post |
| | **_59%_** | 23% | 15% | 3% | Algebra Post |
| 63 | **_58%_** | 4% | 37% | 1% | Upper Post |
| | **_69%_** | 10% | 20% | 1% | Calc Post |
| | **_56%_** | 16% | 26% | 2% | Algebra Post |
| 64 | **_49%_** | 43% | 6% | 2% | Upper Post |
| | **_56%_** | 32% | 10% | 2% | Calc Post |
| | **_59%_** | 29% | 10% | 2% | Algebra Post |
| 65 | **_62%_** | 25% | 12% | 1% | Upper Post |
| | **_70%_** | 18% | 10% | 1% | Calc Post |
| | **_55%_** | 26% | 17% | 2% | Algebra Post |
| 66 | **_82%_** | 6% | 10% | 2% | Upper Post |
| | **_75%_** | 12% | 10% | 3% | Calc Post |
| | **_64%_** | 18% | 15% | 3% | Algebra Post |
| 67 | 15% | **_78%_** | 4% | 3% | Upper Post |
| | 22% | **_59%_** | 15% | 4% | Calc Post |
| | 29% | **_52%_** | 16% | 3% | Algebra Post |
| 68 | 13% | **_69%_** | 13% | 5% | Upper Post |
| | 12% | **_64%_** | 18% | 7% | Calc Post |
| | 15% | **_63%_** | 18% | 4% | Algebra Post |
| 69 | **_45%_** | 37% | 14% | 4% | Upper Post |
| | **_57%_** | 20% | 22% | 1% | Calc Post |
| | **_42%_** | 34% | 21% | 2% | Algebra Post |
| 70 | **_31%_** | 50% | 13% | 6% | Upper Post |
| | **_53%_** | 31% | 14% | 2% | Calc Post |
| | **_38%_** | 43% | 15% | 3% | Algebra Post |
| 71 | **_70%_** | 20% | 8% | 2% | Upper Post |
| | **_73%_** | 20% | 5% | 2% | Calc Post |
| | **_63%_** | 24% | 12% | 1% | Algebra Post |
| 72 | 18% | 16% | **_65%_** | 1% | Upper Post |
| | 11% | 16% | **_70%_** | 2% | Calc Post |
| | 18% | 25% | **_57%_** | 1% | Algebra Post |
| 73 | **_73%_** | 12% | 11% | 3% | Upper Post |
| | **_66%_** | 11% | 19% | 4% | Calc Post |
| | **_53%_** | 24% | 21% | 2% | Algebra Post |
| 74 | **_72%_** | 12% | 13% | 3% | Upper Post |
| | **_66%_** | 9% | 21% | 4% | Calc Post |
| | **_58%_** | 17% | 21% | 4% | Algebra Post |
| 75 | **_78%_** | 5% | 16% | 1% | Upper Post |
| | **_61%_** | 8% | 30% | 1% | Calc Post |
| | **_45%_** | 19% | 35% | 2% | Algebra Post |
| 76 | 17% | **_83%_** | - | - | Upper Post |
| | 20% | **_80%_** | - | - | Calc Post |
| | 33% | **_67%_** | - | - | Algebra Post |
| 77 | 32% | **_68%_** | - | - | Upper Post |
| | 31% | **_69%_** | - | - | Calc Post |
| | 41% | **_59%_** | - | - | Algebra Post |
| 78 | **_83%_** | 17% | - | - | Upper Post |
| | **_83%_** | 17% | - | - | Calc Post |
| | **_66%_** | 34% | - | - | Algebra Post |

# APPENDIX D: STPFASL-LONG SURVEY POINT BISERIAL COEFFICIENT FOR IN-PERSON AND ONLINE ADMINISTRATIONS

Appendix D shows the Point Biserial Coefficient for in-class and online administrations.

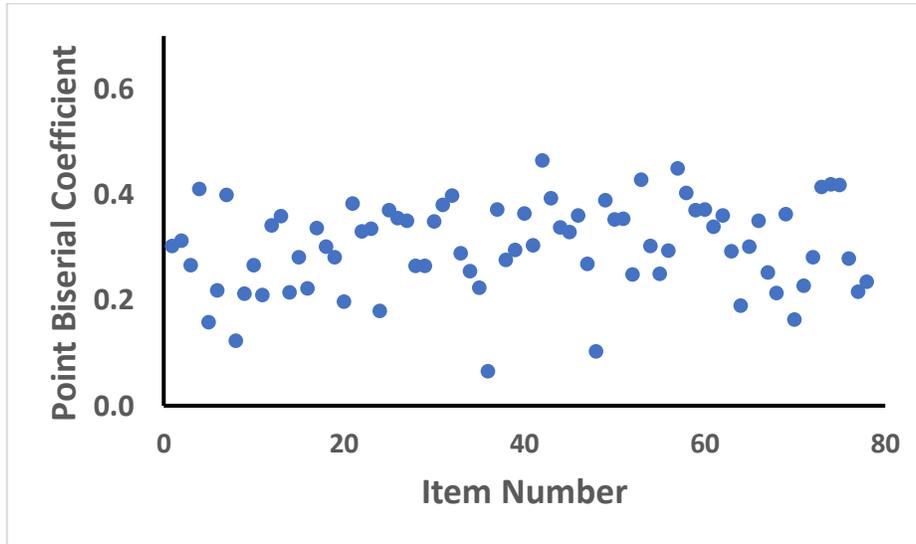

FIGURE 2. Point biserial coefficient, for each item for in-class administration.

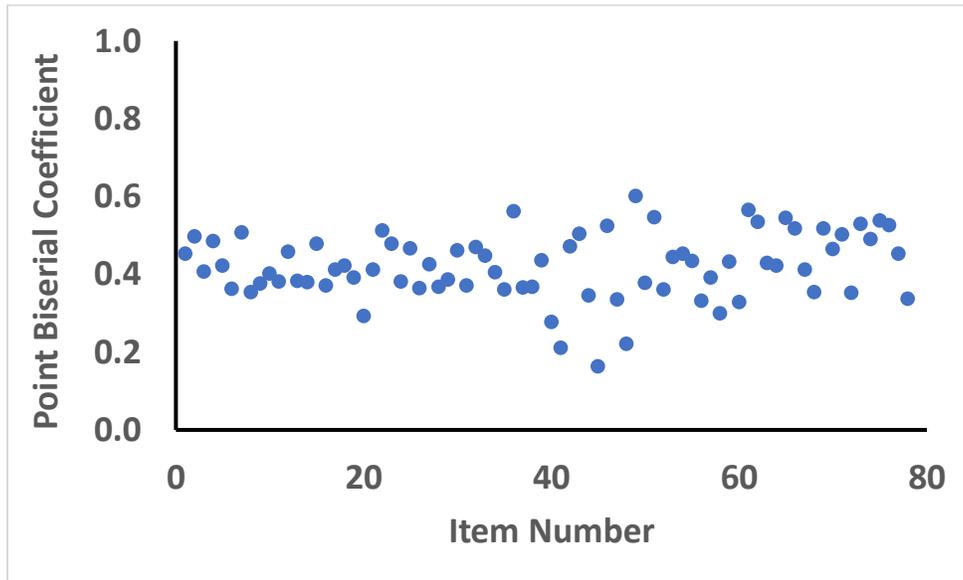

FIGURE 3: Point biserial coefficient for each item for online administration in Qualtrics.